\documentclass[floatfix,%
 reprint,
nofootinbib,
 amsmath,amssymb,
 aps,
]{revtex4-2}

\usepackage{soul} %for striking through text
\usepackage[normalem]{ulem}

\usepackage{graphicx}% Include figure files
\usepackage{dcolumn}% Align table columns on decimal point
\usepackage{bm}% bold math
\usepackage{braket}
\usepackage{amsmath}
\usepackage{amssymb}
\usepackage[hidelinks]{hyperref}
\usepackage[capitalize]{cleveref}
\usepackage{empheq}
\usepackage{tikz}
\usepackage{pgfplots}
\pgfplotsset{compat=1.16}
\usepackage{float}
\usepackage{tikz}
\usepackage{subcaption}
\usepackage{array}
\usetikzlibrary{decorations.markings}
\usetikzlibrary{backgrounds,automata}
\usepackage{comment}
\usepackage{footnote}

\allowdisplaybreaks
\usepackage{prettyref}
\newrefformat{eq}{Eq.~(\ref{#1})}

\begin{document}

\preprint{APS/123-QED}

\title{Microscopic origins of inertial magnetization dynamics} % Force line breaks with \\
\begin{abstract}
    Ultrafast experiments have uncovered inertial magnetization dynamics in ferromagnets, but their microscopic origin remains elusive. Using a non-Markovian quantum master equation we show that inertial dynamics arise from coherent interactions with optical phonons in the lattice. The fast optical frequency explains the nutation observed on picosecond timescales and accounts for variations between experiments through substrate-dependent phonon damping. By establishing magnon–phonon coupling as the microscopic basis of inertial magnetization, our results open new pathways for tailoring ultrafast spin dynamics and controlling magnetic states at terahertz frequencies.
\end{abstract}

\author{Caleb Webb}
\author{Ling Gan}
\author{Shufeng Zhang}%
\affiliation{%
 Department of Physics, University of Arizona\\
}%

\date{\today}% It is always \today, today,
 
\maketitle

Understanding how magnetization responds on ultrafast time scales is increasingly important for both fundamental condensed-matter physics and the development of next-generation spintronic devices. The Landau-Lifshitz-Gilbert equation \cite{gilbert_classics_2004, thiele_momentum_1976} has successfully described magnetization precession and damping across many systems, but recent pump–probe, terahertz, and multi-tone spectroscopy experiments have revealed additional, high-frequency components in the magnetization dynamics: transient nutational oscillations on picosecond and sub-picosecond time scales \cite{neeraj_inertial_2021,li_nutation_2019,de_magnetic_2025,beaurepaire_ultrafast_1996,kimel_ultrafast_2005,stanciu_all-optical_2007,unikandanunni_inertial_2022}. These observations have stimulated renewed theoretical interest in inertial extensions to the LLG equation and in the search for microscopic mechanisms that produce such terms. 

To capture ultrafast magnetization dynamics, extensions of the Landau–Lifshitz–Gilbert equation have been proposed, following reasoning analogous to Gilbert’s original formulation. The so called inertial LLG (iLLG) equation includes a higher order nutational damping term
\begin{align} \label{eq: iLLG}
\frac{d}{dt}\bold{M}=-\gamma\bold{M}\times\bold{H}+\alpha\bold{M}\times\frac{d}{dt}\bold{M}+\eta\bold{M}\times\frac{d^2}{dt^2}\bold{M}.
\end{align}
Where ${\bf M}$ is the magnetization, $\gamma$ is gyromagnetic ratio, ${\bf H}$ is the total magnetic field, $\alpha $ is the damping parameter, and $\eta$ is the nutation (inertial) time. The proposed justification of the inertial term is 
based on a possible misalignment of the non-equilibrium magnetization and angular momentum vectors \cite{wegrowe_magnetization_2012,ciornei_magnetization_2011}. While the resulting equation has captured the dynamics at a short-time scale, there is little evidence that it is supported microscopically, and it remains unclear what physical interaction controls the parameter $\eta$. Further theoretical studies of the microscopic origin of nutation include contributions from the internal angular-momentum structure of many-spin systems \cite{wegrowe_magnetization_2012}, atomistic coarse-graining \cite{bhattacharjee_atomistic_2012,bottcher_significance_2012,bastardis_magnetization_2018}, and non-Markovian interactions with external reservoirs \cite{thibaudeau_emerging_2021,quarenta_bath-induced_2024,suhl_theory_1998}. Crucially, analyses that explicitly retain memory effects show that coupling to a reservoir with finite correlation time generates both Gilbert-type damping and higher-order inertial terms in a systematic expansion of the memory kernel \cite{thibaudeau_emerging_2021,quarenta_bath-induced_2024}. Experimentally, nutation signatures have been reported in a range of systems and with different techniques. Yet observed nutation frequencies and damping vary substantially between studies, even for similar magnetic materials \cite{neeraj_inertial_2021,de_magnetic_2025}, indicating that environmental factors (substrates, interfaces, and phonon dissipation) play a key role. 

In this paper we address these issues by deriving magnetization dynamics from a microscopic magnon-phonon Hamiltonian using a non-Markovian quantum master equation. We demonstrate that allowing memory effects in the reservoir directly leads to inertial (nutational) dynamics in the average magnetization. Similar memory effects have been introduced in \cite{quarenta_bath-induced_2024,thibaudeau_emerging_2021}. While these models are able to describe nutation, they rely on classical treatments of the magnetization vector and cannot give an accurate description of the microscopic origin of the nutation frequency. By using a microscopic description of the spin system we are able to show that the transverse components of the magnetization couple primarily to high frequency optical phonon modes via strain-modulated magnetic anisotropy. If an optical phonon is excited thermally or optically during the excitation protocol it will create an effective spin-spin interaction which is non-local in time and persists for the lifetime of the phonon. Thus nutation can be directly understood as response of the spin system to oscillations in the anisotropy at the optical frequency, typically on the order of a few terahertz. Differences in excitation lifetime and phonon frequency, set by substrate and interface properties for thin films, may provide a natural and readily verifiable explanation to the variability of nutation time scales reported in experiments \cite{neeraj_inertial_2021,de_magnetic_2025}. 

As magnons are the fundamental low-energy excitations of magnets, the change of the energy and angular momentum of a dynamic system must involve the exchange between magnons and other degrees of freedom. For a minimum model of the damping dynamics, we consider a uniformly magnetized ferromagnet whose magnon spectrum is $\omega_{\bf k}$. Within the spin wave approximation the components of the average magnetizaiton per lattice site can be written in terms of the magnon operators as $M_x=-\frac{M_s}{\sqrt{2NS}}\braket{a^{\dagger}_0+a_0}$, $M_y=-\frac{iM_s}{\sqrt{2NS}}\braket{a^{\dagger}_0-a_0}$ and $M_z=-M_s\left(1-\frac{1}{NS}\sum_k\braket{a^{\dagger}_ka_k}\right)$, where $\braket{a_k}=Tr(\rho_m a_k)$ is the quantum expectation value of the annihilation operator $a_k$ with magnon system density matrix $\rho_m$. $M_s=\gamma S$ is the saturation magnetization per spin and $N$ is the total number of spins in the lattice. 

For the magnon-phonon coupling we assume a momentum conserving, bilinear interaction $H_I=\sum_{k}V_{k}a^{\dagger}_kB_k+h.c.$ where the $B_k=b_{-k}^{\dagger}+b_{k}$ is the lattice displacement operator in momentum space and $b_k$ are phonon annihilation operators. Interactions of this form represent the lowest order contributions (within the Holstein-Primakoff expansion) to the magnetoelastic coupling \cite{streib_magnon-phonon_2019,ruckriegel_magnetoelastic_2014,flebus_magnon-polaron_2017}, and have been shown to produce Gilbert damping in multi-spin systems \cite{norambuena_open_2020}. Microscopically couplings at this order originate from dipole-dipole interactions or, for systems with strong spin-orbit coupling, strain modulated magnetic anisotropy \cite{yildirim_anisotropic_1995,moriya_anisotropic_1960}. However, the transverse magnetization is governed by the $k=0$ magnon modes, while the generation of $k\neq0$ modes primarily influences the magnitude of the magnetization \cite{callen_ferromagnetic_1958}. In the supplementary material we show that for short-ranged, non-local spin-spin interactions (including dipole-dipole interactions) the magnon-phonon vertex vanishes $V_k\rightarrow0$ as $k\rightarrow0$. Because the bilinear coupling is diagonal in momentum, we conclude that fluctuations in the anisotropy, $H_{ani}=-K\sum_i(\mathbf{S}_i\cdot\mathbf{r})^2$, is the primary mechanism by which the transverse magnetization couples to the phonon system. Any distortion of the lattice which affects the electron orbitals will necessarily modify the anisotropy. At $k=0$ acoustic phonons produce a uniform shift of the lattice, which does not distort the crystal. We conclude, then, that only optical phonon modes produce a response in the transverse magnetization dynamics. Finally, because the magnon-phonon interaction vertex $V_k$ is proportional to the variation in the anisotropy $\delta K$, it may be significant even in soft magnets. A detailed derivation of the interaction and vertex is presented in the supplementary materials.

The dynamics of the magnon system have both internal and external contributions
\begin{align} \label{eq: akdot}
    \frac{d}{dt}\braket{a_k}=\braket{i[H_m,a_k]}+{\rm Tr} \left(\frac{d\rho_m}{dt}a_k \right).
\end{align}
The first term, with $H_m$ the magnon system Hamiltonian, leads directly to the Larmor precession and the second describes the effect of coupling to the reservoir. Following methods similar to \cite{norambuena_open_2020} we use a Quantum Master Equation (QME) to describe the open system dynamics of the magnons
\begin{align}\label{eq: mastereq}
\frac{d}{dt}\rho_m=- {\rm Tr}\left\{\int_0^t\left[H_I(t),\left[H_I(s),\rho_m(s)\otimes\rho_R\right]\right]ds\right\}.
\end{align}
Where $\rho_m(t)$ and $\rho_R$ are, respectively, the magnon system and phonon reservoir density matrices. The phonon system is assumed to relax on a timescale much faster than the magnon system so that $\rho_R$ remains in approximate equilibrium. In this limit the above equation is obtained by a weak coupling approximation to the Von-Neumann evolution of the magnon system density matrix \cite{breuer_theory_2002,grabert_projection_1982} and is, importantly, \textit{non Markovian}- we allow the evolution of the magnon system at time $t$ to be influenced by its state at earlier times $s<t$. 

The transverse magnetization dynamics can be obtained from \prettyref{eq: akdot} and \prettyref{eq: mastereq}. This is then extrapolated to the full, nonlinear dynamics by noting that, within the spin wave approximation, $M_z^2\approx M_s(2M_z-1)$ and $M_{x/y}M_z\approx M_sM_{x/y}$. After disregarding the eccentricity of the elliptical dynamics (due to the easy axis being misaligned with the applied field) and dropping higher order, non-momentum conserving terms we have, up to the second order in the interaction $V$,
\begin{align}\label{eq: Mdot}
    \dot{\mathbf{M}}=-\gamma\mathbf{M}\times\mathbf{H}_{eff}+\frac{\mathbf{M}\times\mathcal{T}*\mathbf{M}}{M_s}+\frac{\frac{d}{dt}(M^2)\mathbf{M}}{2M_S^2}.
\end{align}
A detailed derivation can be found in the supplementary materials. Above we have introduced the memory kernel
\begin{align}\label{eq: T}
    \mathcal{T}(t-s)=|V_0|^2D_{0}^R(t-s).
\end{align}
$\omega_0=|\gamma| H_0^z+2KS$ is the magnon gap $D_q^R(t-s)=-i\Theta(t-s)\braket{[B_q(t),B_q^{\dagger}(s)]}$ is the retarded phonon Green's function. \prettyref{eq: T} is precisely the magnon self energy within the Born approximation. From \prettyref{eq: Mdot} we can see that the dynamics have been naturally separated into transverse and longitudinal contributions. The last term describes the dependence of $\mathbf{M}$ on the longitudinal dynamics; however, the form of $\frac{d}{dt}M^2$ must be determined independently, and is discussed in the supplementary materials. In this article our focus is on the iLLG equation, which concerns only the transverse dynamics. We will therefore drop the longitudinal dynamics from the following discussion, and comment on their possible significance at the end of the paper.

The resonant frequencies for the magnetization dynamics can be estimated from \prettyref{eq: Mdot} by  assuming a linearized solution $\bold{M}=M_s\hat{z}+\delta\bold{M}$. Taking the Fourier transform of the linearized equation, one finds that the resonances are given by the solutions to $\omega^2-(\mathcal{T}(\omega)-\omega_0)^2=0$. The pole structure of the Green's functions $D^R_0(\omega)$ will introduce new roots at approximately the frequency of the optical mode $\omega_{\pm}\approx\pm\Omega_0$. In the supplementary material it is shown that if the optical mode is broadened (by anharmonic scattering, mass disorder, impurity scattering, etc.) then the Green's function may be decomposed as $D^R_0=D^+-D^-+D^{inc}$.
\begin{align}
    D^{\pm}=\frac{Z}{\omega\mp\Omega_0+iZ\Gamma}
\end{align}
describes the quasiparticle nature of the optical phonons with lifetime $\tau=1/Z\Gamma$. Both the broadening $\Gamma=-Im\Pi_0(\Omega_0)$ and residue $Z=(1-\partial_\omega Re\Pi_0(\Omega_0))^{-1}$ are determined by the phonon self energy $\Pi_0$ evaluated at the poles of the Green's function. $D^{inc}(\omega)$ has no pole structure and accounts for the incoherent background of the reservoir. Far from the poles, $|\omega\pm\Omega_0|>>Z\Gamma$ and $D^R_0(\omega)\sim D^{inc}(\omega)$, suggesting that the spin system is primarily influenced by the incoherent part of the Green's function at low frequencies. Below we show that $D^{inc}$ leads to damping, while the pole at $\omega=\Omega_0$ introduces a high frequency oscillation which we interpret as nutation.

At zero momentum $D^R_0(t)$ is purely real, which imposes parity restrictions on its Fourier transform, namely $ReD^R_0(\omega)$ is even and $ImD^R_0(\omega)$ is odd. From \prettyref{eq: T} we see that the constant shift $D^R(\omega=0)$ is irrelevant to the magnetization dynamics. Thus, at small frequencies we take for the leading contribution to the incoherent Green's function $D^{inc}(\omega)=-i\kappa\omega$, with $\kappa>0$ determined by the sum rule \cite{mahan_many-particle_2000}
\begin{align}
        -\frac{1}{\pi}\int_{-\infty}^{\infty}d\omega ~n_B(\omega)ImD^R_0(\omega)=2n_0+1.
\end{align}
We assume the phonon reservoir to be in thermal equilibrium with inverse temperature $\beta$. $n_B(\omega)$ is the Bose-Einstein distribution and $n_0=n_B(\Omega_0)$ is the number of particles occupying the optical mode. Assuming a frequency cutoff $|\omega|<\Omega_0$ for $D^{inc}$ we find
\begin{align}
        \kappa\approx\frac{2\pi}{\Omega_0^2}(2n_0+1)[1-Z(1+\beta Z\Gamma\delta n_0)].
\end{align}
$\delta n_0=Z\Gamma \partial_{\Omega_0}n_0$ gives the variation in the occupation of the optical mode. For sufficiently small broadening the latter term can be ignored and we find the condition $0<Z<1$, from which we interpret the quasiparticle residue as a measure of the degree of coherence of the reservoir.

Up to quadratic order in the interaction strength we find for the resonant frequencies of the magnetization,
\begin{align} \label{eq: omega}
    \omega\approx\begin{cases}\omega_0+i\kappa\omega_0V^2 \\ \\ \pm\Omega_0-iZ\Gamma-\frac{ZV^2}{\pm\Omega_0-\omega_0-iZ\Gamma} \end{cases}.
\end{align}
The ohmic character of the incoherent contribution to the spectral function, $-ImD^{inc}(\omega)\sim\omega$, leads a Gilbert-type damping with $\alpha=\kappa|V_0|^2/M_s$. This is consistent with a Markovian description of the reservoir. From the Fluctuation Dissipation Theorem the reservoir correlation functions are given by $\braket{B_0(t)B_0(s)}=\frac{1}{\pi}\int d\omega e^{-i\omega (t-s)}n_B(\omega)ImD^R_0(\omega)$. On timescales $\tau_s>>1/\Omega_0$ the spectral function will be Ohmic, and at high temperature the lesser Green's function will be constant, $G^<=n_BImD^R_0\sim1/T$. Thus $\braket{B_0(t)B_0(s)}\sim\delta(t-s)$, suggesting that excitations in the reservoir are uncorrelated on the system timescale. In this white noise limit, the reservoir relaxation time is much faster than that of the system, $\tau_R<<\tau_s$, which is exactly the Markov assumption producing a memoryless dynamics.

Because the transverse components of the magnetization couple only to a single reservoir mode the memory kernel is otherwise featureless apart from the localized resonance at $|\omega-\Omega_0|\lesssim|Z\Gamma|$. This presents the fundamental difference between the predictions of our microscopic model for the spin system and the Caldeira-Leggett model assumed in \cite{quarenta_bath-induced_2024}. In the Calderia-Leggett model the net magnetization is treated as a macroscopic variable which couples to \textit{all} reservoir modes \cite{CL} and broadens the frequency dependence of the memory kernel. While this allows for a small frequency expansion of the kernel and leads directly to the iLLG equation, this procedure is ill suited to capture the high frequency behavior of the magnetization within our framework. Our results are more readily understood from \cref{eq: Mdot,eq: omega}. The coherent part of the Green's function behaves like a damped harmonic oscillator. Defining $\chi(t)\equiv |V_0|^2/M_s(D^+(t)-D^-(t))=(2Z|V_0|^2/M_s)\Theta(t)\sin(\Omega_0t)e^{-Z\Gamma t}$ the transverse components of \prettyref{eq: Mdot} may be rewritten as
\begin{align} \label{eq: eta}
    \dot{\mathbf{M}}=-\gamma\mathbf{M}\times\mathbf{H}_{eff}+\alpha\mathbf{M}\times\dot{\mathbf{M}}+\\
    +\mathbf{M}\times\int_0^t\chi(t-s)\mathbf{M}(s)ds.\nonumber
\end{align}
While this does not quantitatively match the inertial LLG dynamics of \prettyref{eq: iLLG}, it is qualitatively very similar. In the picture presented here the inertial character is understood as being inherited directly from the inertia of the optical vibrations in the lattice. A finite quasiparticle lifetime in the reservoir determines the lifetime of the inertial dynamics. At the scale of the sample magnetization this allows for a coherent interaction between the reservoir and spin system in which energy flows back and forth on a timescale smaller than the phonon thermalization time. On larger timescales the quasiparticles will thermalize instantaneously and the reservoir will be fully decoherent so that the high frequency dynamics can be completely ignored. 

To linear order in deviations from equilibrium \prettyref{eq: iLLG} describes nutation at a frequency $\omega_{\eta}\approx1/\eta$ with broadening $\Delta\omega_{\eta}\approx\alpha/\eta$. This is consistent with \prettyref{eq: eta} if $\eta=1/\Omega_0$ and $Z\Gamma=\alpha\Omega_0$, which is well within the quasiparticle limit and provides the expected scale $\eta\sim 1 ps$. In the iLLG equation the amplitude of the nutation, $|\Delta\theta|$, is fully determined by $\eta$ and $\omega_0$, $|\Delta\theta|_{iLLG}\sim\eta\omega_0$. In \prettyref{eq: eta}, however, the amplitude scales with the additional parameter $|V_0|^2$, and we estimate $|\Delta\theta|\sim|V_0|^2\omega_0/\Omega_0^3$. In order for the weak coupling assumption to hold in \prettyref{eq: mastereq} we require that the system dynamics are not strongly hybridized with the reservoir at the frequency $\omega=\omega_0$, which will be be satisfied for $|V_0|^2<\omega_0\Omega_0$. At the upper limit of the coupling strength, then, the predicted amplitudes scale as $|\Delta\theta|\sim(|\Delta\theta|_{iLLG})^2\sim(\omega_0/\Omega_0)^2$. 

At this limit of the Gilbert damping parameter predicted by our theory is $\alpha\lesssim\frac{\omega_0}{\Omega_0}(1-Z).$ Except at very high temperatures the thermal population $n_0<<1$, and so we have not included this term. Assuming $\omega_0/\Omega_0\sim10^{-3}$, this is at the lower limit for low damping materials, and we interpret it as a correction to full damping parameter which may have many sources \cite{brataas_scattering_2008, gilmore_identification_2007, kambersky_ferromagnetic_1976, hickey_origin_2009}. 

\begin{figure}[tph]
\centering
\includegraphics[width=\columnwidth]{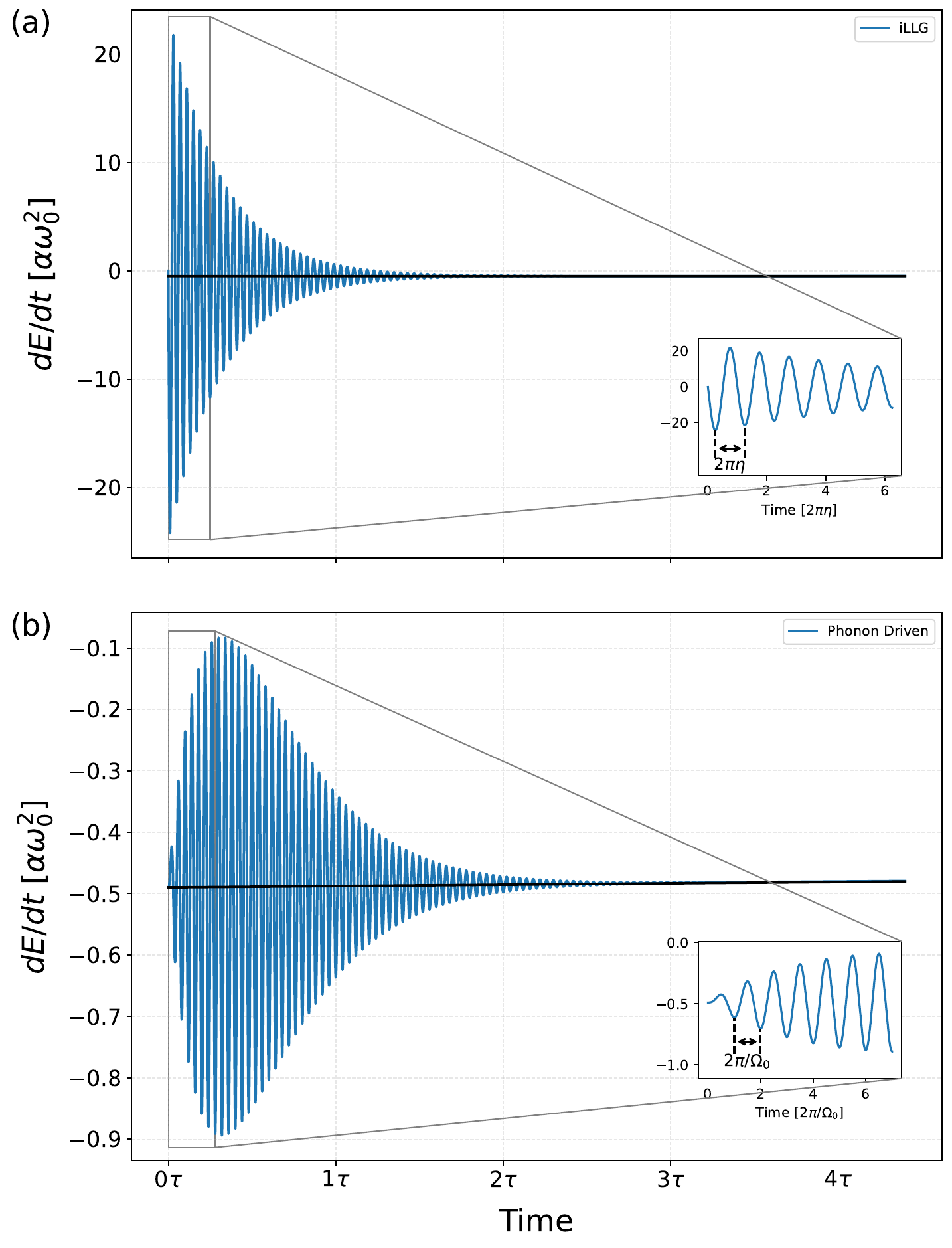}
\caption{\label{Fig1} Rate of magnetization energy fluctuations as described by (a) \prettyref{eq: iLLG} and (b) \prettyref{eq: eta}. The frequencies are chosen relative to the magnon gap $1/\eta\omega_0=\Omega_0/\omega_0=10^3$. The phonon lifetime is chosen to match the damping of the iLLG, $2\tau=1/Z\Gamma=\eta/\alpha$. In both plots $\alpha=0.02$. The black lines in each figure correspond to the non-inertial limits for each equation, $\eta\rightarrow0$ in (a) and $V\rightarrow0$ in (b). The insets show the dynamics on the scale of the high nutation frequency.}
\end{figure}

\begin{figure}[tph]
\centering
\includegraphics[width=\columnwidth]{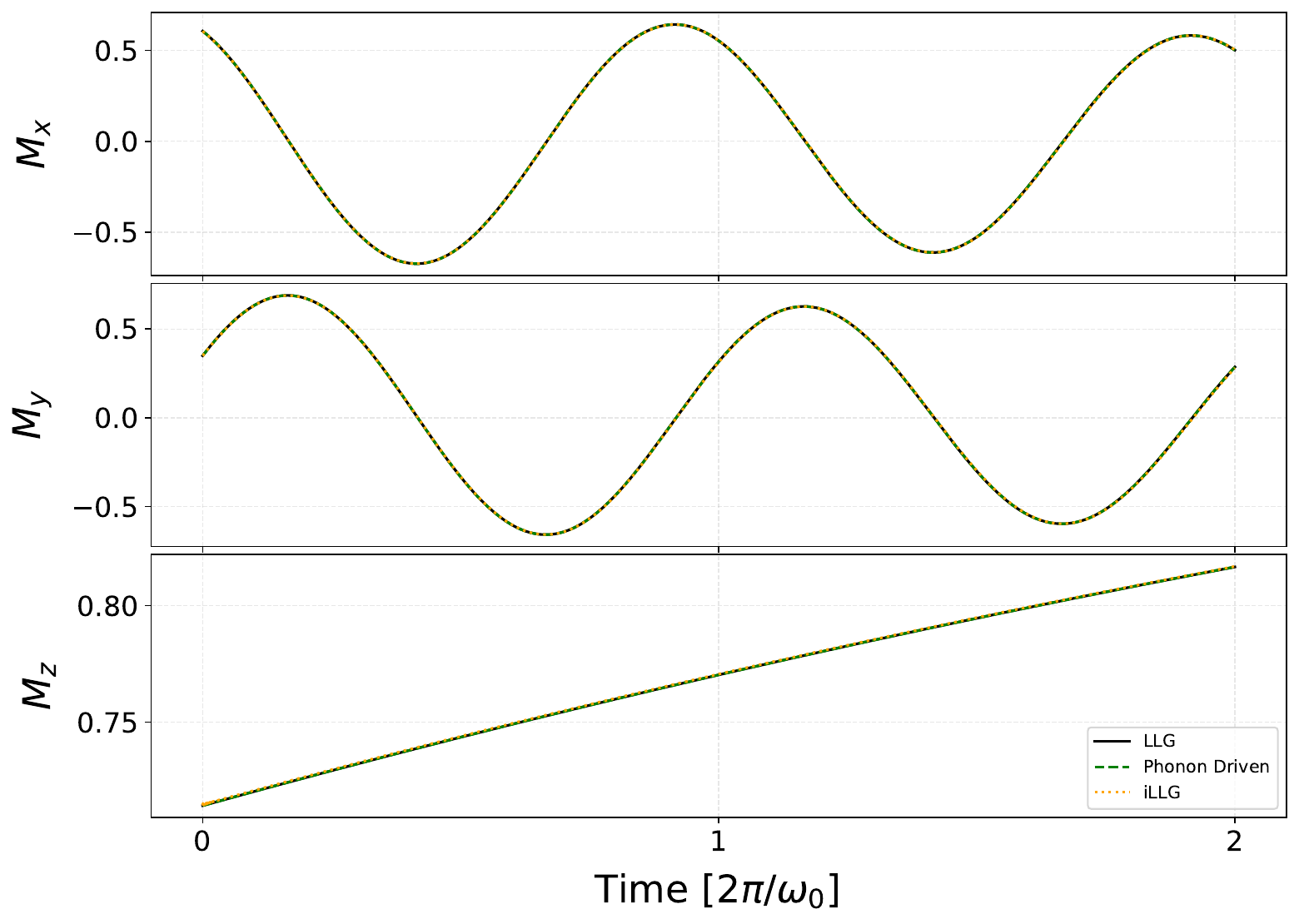}
\caption{\label{Fig2} Magnetization dynamics over two Larmor precessions for \cref{eq: iLLG,eq: eta} and the LLG dynamics. All parameters are chosen as in \cref{Fig1}.}
\end{figure}

To see the comparison more quantitatively, we consider the rate of the change of the magnetic energy, $dE/dt = -{\bf H}_{eff} \cdot d{\bf M}/dt$, as a function of time. Assuming the magnetization
is initially at an angle of $\pi/4$ to an applied magnetic field ${\bf H}$ which is turned on at $t=0$. \cref{Fig1} shows the inertial dynamics over several nutation lifetimes $\tau=1/2Z\Gamma=\eta/2\alpha$ for both the iLLG equation and the phonon driven fluctuations \prettyref{eq: eta}. The parameters are chosen so that the frequency and broadening match in both cases, $\Omega_0=1/\eta$. The insets in each figure show the first few periods of oscillation, from which it can be more clearly seen that the nutation occurs at a frequency $\omega\approx\Omega_0$. In \cref{Fig1}b the amplitude of small fluctuations increases as the spin system accumulates memory over the phonon lifetime $\tau$ before decaying. The iLLG equation, on the other hand, does not describe this initial inertial effect. As expected from the above discussion the amplitude in the iLLG dynamics (\cref{Fig1}a) is much larger than that of the phonon dynamics (\cref{Fig1}b), which results in a larger fluctuation of the magnetization. On longer timescales the inertial effects are damped out and the dynamics align with the LLG equation. \cref{Fig2} shows the Larmor precession for the LLG, iLLG and \prettyref{eq: eta}.

Before we conclude our paper, we wish to comment on several qualitative and quantitative implications of our results.
First, the interaction with the thermal reservoir not only gives rise to the nutation dynamics of the magnetization rotation but also induces longitudinal magnetization dynamics, as shown in the last term of \prettyref{eq: Mdot}. As discussed in the supplementary materials the dynamical equation for $\frac{d}{dt}M^2$ is governed by the $k\neq0$ modes of the magnon system, consistent with \cite{callen_ferromagnetic_1958}. The magnon-phonon interaction introduces an effective coupling  between the magnon modes. Population and subsequent scattering of the higher $k$ modes causes the magnitude of the magnetization to decay over time. This is expected as the thermal reservoir will gradually destroy the long range order in the spin system. Although such longitudinal dynamics are typically negligible well below the Curie temperature $T_c$ where the magnetization amplitude remains nearly constant, they become essential near $T_c$, as ultrafast laser heating experiments can drive demagnetization and remagnetization processes \cite{kirilyuk_ultrafast_2010,beaurepaire_ultrafast_1996}. %[\textcolor{red}{Ref-a}]. 
Previous phenomenological models \cite{garanin_fokker-planck_1997,xu_magnetization_2012} %[\textcolor{red}{Ref-b}] 
have incorporated longitudinal relaxation phenomenologically; our work provides a microscopic foundation for these models, as explicitly reflected in the last term of \prettyref{eq: Mdot}

The present formalism can be naturally extended to antiferromagnets. Because the antiferromagnetic resonance frequency is several orders of magnitude higher than the ferromagnetic Larmor frequency which is comparable to the optical phonon frequency, care must be taken in determining the high frequency contribution to the memory kernel, since the $k=0$ mode may strongly hybridize with the optical modes in the reservoir. A consistent treatment of inertial dynamics in antiferromagnets will therefore require further theoretical development.

Finally, the correlation between nutation dynamics and reservoir lifetimes is not restricted to phonons. In ferromagnet/metal heterostructures, magnon–electron scattering dominates magnetic relaxation and produces the well-known conductivity- and resistivity-like damping via intra- and inter-band transitions \cite{gilmore_identification_2007,fahnle_generalized_2011,suhl_theory_1998}. When conduction electrons are treated as a thermal reservoir, the nutational response is governed by the electron spin-flip scattering time—typically much shorter than a picosecond. Consequently, while adjacent metallic layers enhance the overall damping parameter, they have little influence on the nutation dynamics, which remain primarily determined by phonon relaxation.

The work was supported by the National Science Foundation (Award numbers 2401267 and 2425567).

\bibliography{iLLG_refs}
\end{document}

\preprint{APS/123-QED}

\title{\textcolor{red}{\hsout{Colossal} Large} Magnon Magnetoresistance of Two-Dimensional \textcolor{red}{Ferrom}agnets}% Force line breaks with \\

\author{Caleb Webb}
\author{Shufeng Zhang}%
\affiliation{%
 Department of Physics, University of Arizona\\
}%

\date{\today}% It is always \today, today,
 
\begin{onecolumngrid}
\section{Magnon Picture}
The magnetization of a sample is a classical, macroscopic vector which, however, emerges from the individual, quantum spins on each site. The dynamics of the classical magnetization are therefore governed by the microscopic spin dynamics, which are themselves most easily described using the Holstein-Primakoff (HP) transformation. Within the spin wave approximation the cartesian components of the net magnitization per site can be written as
\begin{align*}
M_x=-\frac{\gamma}{N}\braket{\SUM{i}{}S_i^x}&=-\frac{\gamma}{N}\sqrt{\frac{S}{2}}\SUM{i}{}\braket{a_i^{\dagger}+a_i}\\
&=-\gamma\sqrt{\frac{S}{2N}}\braket{a^{\dagger}_{k=0}+a_{k=0}}.
\end{align*}
Above $\gamma$ is the gyromagnetic ratio and we have used the Fourier transform $a_i=1/\sqrt{N}\sum_{k}e^{ikR_i}a_k$ and $\delta_{k,k'}=1/N\sum_ie^{-i(k-k')R_i}$, with $N$ the total number of lattice sites. The brakets represent the quantum expectation value $\braket{A}=Tr(\rho_mA)$, with $\rho_m$ the density matrix for the magnon system. Defining the saturization magnetization per site as $M_S=-\gamma S$ we may write the magnetization components as
\begin{align}
&M_x/M_S=\frac{1}{\sqrt{2NS}}\braket{a^{\dagger}_0+a_0}\\
&M_y/M_S=\frac{i}{\sqrt{2NS}}\braket{a^{\dagger}_0-a_0}\\
&M_z/M_S=1-\frac{1}{NS}\SUM{k}{}\braket{a^{\dagger}_ka_k}.
\end{align}
In applying the HP transformation we have assumed a quantization axis along the $\hat{z}-$direction. Notably, the transverse components depend only on the $k=0$ magnon, which describes a uniform tilting of the magnetic moments away from the $\hat{z}-$axis. The longitudinal component depends on all $k$ states, but are higher order in the magnon operators and perturbative parameter $\frac{1}{\sqrt{NS}}$. It can be shown that the longitudinal dynamics are described purely by the $k\neq0$ modes \cite{callen_ferromagnetic_1958} and fluctuations in the variance of the net magnetization $\sigma^2_M=(\braket{M^2}-\braket{M}^2)/M_S^2$. To quadratic order in the magnon operators
\begin{align}
    \sigma^2_M&=\frac{1}{N^2S^2}\SUM{i,j}{}\left(\braket{\mathbf{S}_i\cdot\mathbf{S}_j}-\braket{\mathbf{S}_i}\cdot\braket{\mathbf{S}_j}\right)\nonumber\\
    &=\frac{1}{N^3S}\SUM{i,j}{}\SUM{k,k'}{}e^{-i(\mathbf{R}_i\cdot\mathbf{k}-\mathbf{R}_j\cdot\mathbf{k'})}\left(\braket{a_ka^{\dagger}_{k'}+a^{\dagger}_{k'}a_k}-2\braket{a^{\dagger}_{k'}}\braket{a_k}\right)\nonumber\\
    &=\frac{1}{NS}\left(\braket{a_0a^{\dagger}_0+a^{\dagger}_0a_0}-2\braket{a^{\dagger}_0}\braket{a_0}\right)\nonumber\\ \nonumber\\
    \implies\sigma^2_M&=\frac{1}{NS}+\frac{2}{NS}\left(\braket{a^{\dagger}_0a_0}-\braket{a^{\dagger}_0}\braket{a_0}\right).\label{eq:sigM}
\end{align}
In the Ferromagnetic ground state the variance $\sigma_M^2=1/NS$, which we understand as the constant contribution in \cref{eq:sigM}. The second term describes an additional broadening of the variance by fluctuations in the $k=0$ mode. This contribution is maximized in equilibrium, where the magnon operators are uncorrelated ($\braket{a^{\dagger}}=\braket{a}=0$). The magnitude of the magnetization is
\begin{align*}
M^2/M_S^2=1-\frac{2}{NS}\left(\braket{a_0^{\dagger}a_0}-\braket{a_0^{\dagger}}\braket{a_0}\right)-\frac{2}{NS}\SUM{k\neq0}{}\braket{a_k^{\dagger}a_k},
\end{align*}
Where the correlation in the $k=0$ magnon operators has appeared because the transverse and longitudinal components of the magnetization are of different orders in the HP expansion. From \prettyref{eq:sigM} one finds that this can be written in terms of the variance as
\begin{align}\label{eq: Msquared}
    M^2/M_S^2=1+\frac{1}{NS}-\sigma^2_M-\frac{2}{NS}\SUM{k\neq0}{}\braket{a_k^{\dagger}a_k}.
\end{align}

Within the spin wave approximation the longitudinal component of the magnetization can be written in terms of the magnitude. $M_z^2\approx M_S(2M_z-1)$, and so
\begin{align}\label{eq: Mz}
    M_z\approx\frac{1}{2M_S}\of{1+M^2-M_x^2-M_y^2}.
\end{align}
The dynamics, then, can be fully described by the evolution of $\braket{a_0}$ and $\braket{a^{\dagger}_ka_k}$ for $k\neq0$.

\subsection{Magnon System Hamiltonian}
In \prettyref{section: S2A} we show that any spin Hamiltonian producing a bilinear coupling between the transverse spin components and the phonon reservoir must contain single ion anisotropy. Supposing an easy-axis anisotropy along the $\hat{r}$ direction we then have
\begin{align}
    \hat{H}=-J\sum_{\braket{i,j}}\mathbf{S}_i\cdot\mathbf{S}_j-K\sum_i(\mathbf{S}_i\cdot\hat{r})^2-\gamma\sum_i\mathbf{H}\cdot\mathbf{S}_i.
\end{align}
The anisotropy can be viewed as a tensor $K_{\alpha\beta}=Kr_{\alpha}r_{\beta}$, with $\hat{r}=(x,y,z)^T$. Applying the HP transformation one obtains, up to second order in the magnon operators, $\hat{H}_{m}=\hat{H}^{(0)}+\hat{H}^{(1)}+\hat{H}^{(2)}$ with
\begin{align}
    &\hat{H}^{(0)}=-JS^2NN_{\delta}-\gamma H^zSN-KS^2Nz^2 \label{eq: H0}\\
    &\hat{H}^{(1)}=-\sqrt{\frac{S}{2}}\sum_i\left\{\left[\gamma(H^x+iH^y)+2KSz(x+iy)\right]a^{\dagger}_i+h.c.\right\}\label{eq: H1}\\
    &\hat{H}^{(2)}=\sum_k\left\{\omega_ka_k^{\dagger}a_k-\left[\frac{S}{2}K(x+iy)^2a_k^{\dagger}a_{-k}^{\dagger}+h.c.\right]\right\}.\label{eq: H2}
\end{align}
Where we have applied the Fourier transform $a_i=\frac{1}{\sqrt{N}}\sum_ke^{i\mathbf{k}\cdot\mathbf{R}_i}a_k$. Above $N_{\delta}$ is the number of nearest neighbors to a given site, and $\omega_k=2JSN_{\delta}(1-\gamma_k)+KS(3z^2-1)+\gamma H^z$ is the usual magnon dispersion relation with structure factor $\gamma_k=\frac{1}{N_{\delta}}\sum_{\delta}e^{i\mathbf{k}\cdot\mathbf{\delta}}$.

The equilibrium magnetization direction is that which minimizes $\hat{H}^{(0)}$, and lies between $\hat{r}$ and the external field $\mathbf{H}$. In order for the spin system to relax towards the classical equilibrium, we choose the quantization axis for the spin directions such that $S_z$ lies along the $\mathbf{M}_{eq}$, see \prettyref{fig: coords}. 
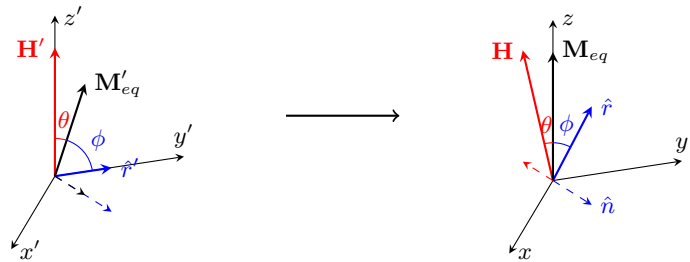
\begin{figure}[th]
    \centering
\begin{tikzpicture}[x=1.5cm,y=1.7cm,z=1cm,>=stealth,rotate around y=-15]
% The axes

\def\th{30}
\def\thr{70}
\def\ph{45}

\draw[->] (xyz cs:z=0) -- (xyz cs:z=-1) node[right] {$x'$};
\draw[->] (xyz cs:y=0) -- (xyz cs:y=1.25) node[right] {$z'$};
\draw[->] (xyz cs:x=0) -- (xyz cs:x=1) node[above] {$y'$};

\node (O) at (0,0,0) {};

\draw[->,red,thick] (0,0,0) -- (0,1,0) coordinate (H) node[left] {$\mathbf{H}'$};

\draw[->,blue,thick] (0,0,0) -- ({sin(\thr)*sin(\ph)},{cos(\thr)},{-sin(\thr)*cos(\ph)}) coordinate (r) node[right] {$\hat{r}'$};

\draw[->,thick] (0,0,0) -- ({sin(\th)*sin(\ph)},{cos(\th)},{-sin(\th)*cos(\ph)}) coordinate (S) node[right] {$\mathbf{M}_{eq}'$};

\path (H) -- (O) -- (S)
    pic ["$\theta$", draw, red, angle eccentricity=1.5] {angle = S--O--H};

\path (S) --  (O) -- (r)
    pic ["$\phi$", draw, blue, angle eccentricity=1.5] {angle = r--O--S};

\draw[dashed,blue,->] (0,0,0) -- ({sin(\thr)*sin(\ph)},0,{-sin(\thr)*cos(\ph)});
\draw[dashed,black,->] (0,0,0) -- ({sin(\th)*sin(\ph)},0,{-sin(\th)*cos(\ph)});

\end{tikzpicture}
\begin{tikzpicture}
    \path (0,2) -- (0,-2);
    \draw[thick,->] (0,0) -- (1.5,0);
    \path (-1,0) -- (0,0);
        \path (1.5,0) -- (2.5,0);
\end{tikzpicture}
\begin{tikzpicture}
[x=1.5cm,y=1.7cm,z=1cm,>=stealth,rotate around y=-15]
% The axes

\def\th{30}
\def\thr{40}
\def\ph{45}

\draw[->] (xyz cs:z=0) -- (xyz cs:z=-1) node[right] {$x$};
\draw[->] (xyz cs:y=0) -- (xyz cs:y=1.25) node[right] {$z$};
\draw[->] (xyz cs:x=0) -- (xyz cs:x=1) node[above] {$y$};

\node (O) at (0,0,0) {};

\draw[->,red,thick] (0,0,0) -- ({-sin(\th)*sin(\ph)},{cos(\th)},{sin(\th)*cos(\ph)}) coordinate (H) node[left] {$\mathbf{H}$};

\draw[->,blue,thick] (0,0,0) -- ({sin(\thr)*sin(\ph)},{cos(\thr)},{-sin(\thr)*cos(\ph)}) coordinate (r) node[right] {$\hat{r}$};

\draw[->,thick] (0,0,0) -- (0,1,0) coordinate (S) node[right] {$\mathbf{M}_{eq}$};

\path (H) -- (O) -- (S)
    pic ["$\theta$", draw, red, angle eccentricity=1.5] {angle = S--O--H};

\path (S) --  (O) -- (r)
    pic ["$\phi$", draw, blue, angle eccentricity=1.5] {angle = r--O--S};

\draw[dashed,blue,->] (0,0,0) -- ({sin(\thr)*sin(\ph)},0,{-sin(\thr)*cos(\ph)}) node[right] (n) {$\hat{n}$};
\draw[dashed,red,->] (0,0,0) -- ({-sin(\th)*sin(\ph)},0,{sin(\th)*cos(\ph)});

\end{tikzpicture}
\caption{\label{fig: coords} Quantization axes are chosen so that the $z$-component of the spin operators lies along the equilibrium magnetization direction $\mathbf{M}_{eq}$. The angle $\theta$ between $\mathbf{M}_{eq}$ and $\mathbf{H}$ is determined by minimizing \prettyref{eq: H0}, or equivalently setting $\hat{H}^{(1)}=0$.}
\end{figure}
Minimizing \prettyref{eq: H0} with respect to $\theta$ yields $\gamma H\sin(\theta)=2KS\sin(\phi)\cos(\phi)$. As described in the following section the expansion of the anisotropy about its local equilibrium in the crystal field $K\approx K(\mathbf{R}_i^{(0)})+\nabla K\cdot(\mathbf{R}_i-\mathbf{R}_i^{(0)})$ produces the bilinear magnon-phonon coupling used in the main text. Applying the constraint in $\theta$ to \prettyref{eq: H1} we see that the equilibrium contribution to $\hat{H}^{(1)}$ also vanishes, leaving only the interaction (see also \prettyref{eq: vertex}).
\begin{align}
    \hat{H}^{(1)}=\hat{H}_I=-(2S)^{3/2}z\sum_i\left[\frac{x+iy}{2}\nabla K\cdot\delta\mathbf{R}_ia_i^{\dagger}+h.c.\right].
\end{align}
In an isolated system the magnon operators will evolve according to $\hat{H}_m=\hat{H}^{(2)}$. The Bogoliubov terms in \prettyref{eq: H2} complicate the transverse dynamics slightly since $i[\hat{H}^{(2)},a_p]=-i\omega_pa_p+iKS(x+iy)^2a_{-p}^{\dagger}$. For dynamics of the transverse magnetization we obtain
\begin{align} \label{eq: transverse_isolated}
    \frac{d}{dt}\begin{pmatrix} M_x \\ M_y\end{pmatrix} = \begin{pmatrix} 0 & \omega_0\\ -\omega_0 & 0 \end{pmatrix}\begin{pmatrix} M_x \\ M_y \end{pmatrix} + KS\begin{pmatrix} -2xy & x^2-y^2 \\ x^2-y^2 & 2xy \end{pmatrix} \begin{pmatrix} M_x \\ M_y \end{pmatrix}.
\end{align}
The matrix in the second term above we recognize as twice the symmetric part of the matrix $W=\mathbf{n}\mathbf{n}^TX$ with vector $\mathbf{n}=(x,y)^T$ and matrix $X=\begin{pmatrix} 0 & 1 \\ -1 & 0 \end{pmatrix}$. The antisymmetric part is $W_A=n^2X/2$. Using $X\mathbf{M}=\mathbf{M}\times\hat{z}$ we can write \prettyref{eq: transverse_isolated} as 
\begin{align}\label{eq: ellipse}
    \dot{\mathbf{M}}=(\omega_0-KSn^2)\mathbf{M}\times\hat{z}+2KSn^2\hat{\mathbf{n}}\cdot(\mathbf{M}\times\hat{z})\hat{\mathbf{n}}.
\end{align}
The above describes elliptical precession with frequency $\omega=\sqrt{\omega_0^2-(KSn^2)^2}$ about the $\hat{z}$ axis with semi-major axis $\hat{n}$ in the $xy$-plane as shown in \prettyref{fig: ellipses}. The semi-major axis has unit length, and the semi-minor axis has length $b=\sqrt{\frac{\omega_0-KSn^2}{\omega_0+KSn^2}}$. For a strong external field $\omega_0>>KS$ and so the eccentricity of the ellipse is $\epsilon\approx n\sqrt{\frac{2KS}{\omega_0}}<<1$. Thus, for an external field misaligned with the easy axis one should expect an \textit{elliptical} precession about the effective field direction. The eccentricity of the orbit will be small, however, so that the dynamics can be treated as approximately circular.

\begin{figure}
    \centering
    \begin{subfigure}[t]{0.45\textwidth}
    \centering
    \begin{tikzpicture}
    [tdplot_main_coords]
    \draw[->] (0,0,0) -- (0,0,2.25) node[right] {$z$};
    \draw[->] (0,0,0) -- (0,1.75,0) node[right] {$y$};
    \draw[->] (0,0,0) -- (2.25,0,0) node[right] {$x$};
    \draw[thick,red] plot[domain=0:2*pi,samples=100,smooth] 
        ( {cos(\x r)*cos(45)-sin(\x r)/(1.5)*sin(45)}, {-cos(\x r)*sin(45)-sin(\x r)/(1.5)*cos(45)}, 1.5 );
    \draw[thick,-latex] (0,0,0) -- ( {cos(-3*pi/4 r)*cos(45)-sin(-3*pi/4 r)/(1.5)*sin(45)}, {-cos(-3*pi/4 r)*sin(45)-sin(-3*pi/4 r)/(1.5)*cos(45)}, 1.5 ) node[right] {$\mathbf{M}$};
    \end{tikzpicture}
    \caption{Magnetization precesses elliptically about the $z$-axis.}
    \end{subfigure}
    ~
    \begin{subfigure}[t]{0.45\textwidth}
    \centering
    \begin{tikzpicture}[scale=1.25]
    \draw[->] (0,0) -- (0,1.5) node[right] {$y$};
    \draw[->] (0,0) -- (1.5,0) node[right] {$x$};
    \draw[thick,red] plot[domain=0:2*pi,samples=100,smooth] 
        ( {cos(\x r)*cos(45)-sin(\x r)/(1.25)*sin(45)}, {-cos(\x r)*sin(45)-sin(\x r)/(1.25)*cos(45)} );
    \draw[thick,blue,dashed,->] (0,0) --node[anchor=190] {$\hat{n}$} ( {cos(45)}, {-sin(45)} );
    \draw[thick,blue,dashed,->] (0,0) --node[anchor=310  ] {$b\hat{n}_{\perp}$} ({-1/(1.25)*sin(45)}, {-1/(1.25)*cos(45)} );
    \end{tikzpicture}
    \caption{Elliptical orbit of the magnetization in the $xy$-plane. The vector $\hat{n}$ defines the semi-major axis with unit length. The semi-minor axis has length $b=\sqrt{\frac{\omega_0-KSn^2}{\omega_0+KSn^2}}$.}
    \end{subfigure}
    \caption{\label{fig: ellipses}}
\end{figure}

\section{Open System Dynamics}
Our aim is to determine the dynamics of the total magnetization and, ultimately, to derive a microscopic justification for the LLG equation. As described in the previous section, for an isolated spin system under the influence of an effective field $\mathbf{H}$ one should expect a lossless dynamics- i.e. simple Larmor precession $\frac{d}{dt}\mathbf{M}=-\gamma\mathbf{M}\times\mathbf{H}$, with $\gamma$ the gyromagnetic ratio. The $LLG$ equation includes a phenomenological dissipation term $\frac{d}{dt}\mathbf{M}=-\gamma\mathbf{M}\times\mathbf{H}+\alpha\mathbf{M}\times\frac{d}{dt}\mathbf{M}$, which can be understood microscopically if the spin system is \textit{not} isolated from its environment. Angular momentum must be globally conserved within the Ferromagnetic sample, and any loss in total magnetization must be transferred to the lattice. The dissipation in the LLG equation may therefore be described microscopically by a spin-phonon interaction, which we model as the simple magnon-phonon interaction
\begin{align}  \label{eq: HI}
\hat{H}_I=\SUM{k}{}(V_{k}a_k^{\dagger}B_k+h.c.).
\end{align}
Above, $a^{\dagger}_k$ creates a magnon with momentum $k$ and $B_k=b_k+b_{-k}^{\dagger}$ is the lattice displacement operator in momentum space- $b^{\dagger}_k$ creates a phonon with momentum $k$. 

\subsection{Magnon-Phonon Scattering}\label{section: S2A}
The bi-linear interaction \prettyref{eq: HI} represent the lowest order contributions to the magnetoelastic scattering \cite{streib_magnon-phonon_2019,ruckriegel_magnetoelastic_2014,flebus_magnon-polaron_2017}, which originate from dipole-dipole interactions and strain-modulated magnetic anisotropy in materials with strong spin-orbit coupling \cite{yildirim_anisotropic_1995,moriya_anisotropic_1960}. The vertex is diagonal due to translational invariance of the interaction, assuming that either $V(\mathbf{r},\mathbf{r'})=V(\mathbf{r}-\mathbf{r'})$ is nonlocal or $V(\mathbf{r},\mathbf{r'})=V$ is constant. The transverse dynamics of the spin system will therefore be governed by the long-wavelength ($\mathbf{k}\rightarrow0$) behavior of $V_k$. In this section we show that the vertex $V_k$ vanishes at zero momentum for any non-local spin-spin interaction which is not sufficiently long range, $V\sim|\mathbf{r}-\mathbf{r'}|^{-2}$.

Consider a non-local spin-spin interaction  $V=\sum_{i,j}V_{\alpha\beta}(\mathbf{R_i}-\mathbf{R_j})S^{\alpha}_iS^{\beta}_j$, with $\alpha$ and $\beta$ summing over cartesian spin components. We apply the usual expansion in small ion displacement $\mathbf{R_i}=\mathbf{R_i}^0+\mathbf{u}_i$. Defining the difference $\mathbf{R_i}^0-\mathbf{R_j}^0=\mathbf{\delta}$ and expanding $V$ up to linear order in $\mathbf{u}_i$
\begin{align} \label{eq: expanded}
    V=V^0+\sum_{i,\delta}\nabla V_{\alpha\beta}(\mathbf{\delta})\cdot(\mathbf{u}_i-\mathbf{u}_{i+\delta})S_i^{\alpha}S^{\beta}_{i+\delta}.
\end{align}
The Fourier transform of the quantized lattice displacements are expressed in terms of the phonon operators $\mathbf{u}_i=\frac{1}{\sqrt{N}}\sum_{q,\lambda}\mathbf{d}_{q,\lambda}e^{i\mathbf{q}\cdot\mathbf{R}_i}(b_{q\lambda}+b_{-q,\lambda}^{\dagger})$. The factor $\mathbf{d}_{q,\lambda}=\frac{\mathbf{e}_{q,\lambda}}{\sqrt{2m\Omega_{q,\lambda}}}$, with $\mathbf{e}_{q,\lambda}$ the polarization vector for phonons with polarization $\lambda$, $m$ the ion mass and $\Omega_{q,\lambda}$ the phonon dispersion relation. The spin operators may be transformed as $S_i^{\alpha}=\frac{1}{\sqrt{N}}\sum_ke^{i\mathbf{k}\cdot\mathbf{R}_i}S_k^{\alpha}$. The phonon contribution to the interaction is therefore
\begin{align}
    \Delta V=\frac{1}{\sqrt{N}}\sum_{k,q}\sum_{\delta,\lambda}\nabla V_{\alpha\beta}\cdot\mathbf{d}_{q,\lambda}(1-e^{i\mathbf{q}\cdot\mathbf{\delta}})e^{-i(\mathbf{k}+\mathbf{q})\cdot\delta}(b_{q,\lambda}+b_{-q,\lambda}^{\dagger})S_k^{\alpha}S_{-k-q}^{\beta}
\end{align}
The difference vectors $\delta$ extend to every point on the lattice, and $V(\delta)$ has a non-trivial Fourier transform, $V(\mathbf{p})=\frac{1}{\sqrt{N}}\sum_{\delta}e^{-i\mathbf{p}\cdot\mathbf{\delta}}V(\mathbf{\delta})$. The difference $\mathbf{u}_i-\mathbf{u}_{i+\delta}$ appearing in \prettyref{eq: expanded} directly leads to a linear dependence on the phonon momentum, since $\frac{1}{\sqrt{N}}\sum_{\delta}\nabla V(1-e^{i\mathbf{q}\cdot\mathbf{\delta}})e^{-i(\mathbf{k}+\mathbf{q})\cdot\delta}=i(\mathbf{k}+\mathbf{q})V(\mathbf{k}+\mathbf{q})-i\mathbf{k}V(\mathbf{k}).$ Here we have assumed that the spin-spin interaction is symmetric, so that $V(-\mathbf{k})=V(\mathbf{k}).$ Finally, we find for the phonon-modulated spin-spin interaction
\begin{align} \label{eq: dV}
    \Delta V=i\sum_{k,q}\sum_{\lambda}[(\mathbf{k}+\mathbf{q})V_{\alpha\beta}(\mathbf{k}+\mathbf{q})-\mathbf{k}V(\mathbf{k})]\cdot\mathbf{d}_{q\lambda}(b_{q,\lambda}+b^{\dagger}_{-q,\lambda})S_k^{\alpha}S_{-k-q}^{\beta}.
\end{align}

The off-diagonal components of $V_{\alpha\beta}$ will yield a coupling of the form \prettyref{eq: HI}. For example, using the HP transformation $S_k^xS^z_{-k-q}+S_k^zS_{-k-q}^x\approx\sqrt{2NS}S[(a^{\dagger}_{-k}+a_k)\delta_{k,-q}+(a^{\dagger}_{k+q}+a_{-k-q})\delta_{k,0}]$. Then from \prettyref{eq: dV}
\begin{align}
    &\hat{V}_{mp}=\sum_{k}W_{mp}(\mathbf{k})a^{\dagger}_kB_k+h.c.\quad\textrm{with vertex}\\
    &W_{mp}(\mathbf{k})=2iS\sqrt{2NS}\sum_{\lambda}\mathbf{k}\cdot\mathbf{d}_{k\lambda}V_{xz}(\mathbf{k}).    
\end{align}
For functions $V(\delta)\sim|\delta|^{-n}$ in $D$-dimensions, the Fourier transform decays as $V(\mathbf{q})\sim|\mathbf{q}|^{n-D}$ and so $\mathbf{q}V(\mathbf{q})\rightarrow0$ as $\mathbf{q}\rightarrow0$ for $n\geq3$. Consequently, phonons do not couple directly to the transverse spin dynamics via the dipole-dipole interaction. Any short range interaction such as the exchange anisotropy will similarly fail to produce magnon-phonon coupling at zero momentum. 

We therefore consider local spin-spin interactions as a source of magnon-phonon coupling at zero momentum. Consider the single ion anisotropy \cite{white_quantum_1983}
\begin{align}
    H_{ani}=-\sum_iK(\mathbf{R}_i)_{\alpha\beta}S_i^{\alpha}S_i^{\beta}.
\end{align}
The anisotropy $K$ is determined by the local crystal field surrounding an ion at $\mathbf{R}_i$. Considering again a small displacement from equilibrium
\begin{align}
    \Delta H_{ani}=\sum_i\nabla K_{\alpha\beta}\cdot\mathbf{u}_iS_i^{\alpha}S_i^{\beta}.
\end{align}
$K$ may vary within the immediate neighborhood of $\mathbf{R}_i^0$, but is otherwise constant throughout the lattice. That is, $K(\mathbf{R}_i^0)=K(\mathbf{R}_j^0)$ for all $i,j$ due to the periodicity of the lattice\footnote{Technically $K$ may vary for different ions within the unit cell, but this does not influence our conclusion in any way.}. $\nabla K(\mathbf{R}_i^0)$ is therefore a constant and so
\begin{align}
    \Delta H_{ani}=\frac{1}{\sqrt{N}}\sum_{k,q,\lambda}\nabla K_{\alpha\beta}\cdot\mathbf{d}_{q,\lambda}B_{q,\lambda}S^{\alpha}_kS^{\beta}_{-k-q}.
\end{align}
The off-diagonal terms again contribute a bilinear magnon-phonon interaction. Define $K_{\pm}=(K_{xz}\pm iK_{yz})/2$, then
\begin{align}\label{eq: vertex}
\hat{H}_I=(2S)^{3/2}\sum_{k,\lambda}\big\{\nabla K_-\cdot\mathbf{d}_{k,\lambda}^*B_{k,\lambda}^{\dagger}a_k+h.c.\big\}
\end{align}
Importantly, this vertex does not vanish in the zero momentum limit. Thus, we conclude that strain modulated single-ion anisotropy provides a mechanism for phonons to couple to the transverse magnetization dynamics.

For itinerant magnets such as CoFeB and Permalloy some care must be taken in interpreting this result. Here the anisotropy emerges as an effective field due to spin orbit coupling mixing the electron orbitals \cite{bruno_tight-binding_1989,katsnelson_first-principles_2000}. $K=K(\{\mathbf{R}_i\})$ is therefore a global parameter which depends on all lattice positions. Any distortion of the lattice which affects the electron orbitals will necessarily modify the anisotropy, and so we have $\nabla K\rightarrow\delta K/\delta\mathbf{R}_i$. 

One may expect that this coupling is stronger for systems with large anisotropy, such as in thin Co films. Because the coupling depends on the variation of the anisotropy it may still be non-negligible in soft magnets. It has been suggested, for example, that a reduction in anisotropy due to magnetoelastic coupling is required to explain the low coercivity in different compositions of $\text{Ni}_x\text{Fe}_{100-x}$ \cite{renuka_balakrishna_solution_2021, lewis_permalloy_1964}. This could serve as motivation that the variation in $K$ with lattice distortion is non-negligible. Temperature and thickness dependent changes in the perpendicular anisotropy of thin CoFeB films has similarly been attributed to the influence magnetoelastic effects \cite{lee_temperature_2017,gowtham_thickness-dependent_2016}. 

\subsection{Master Equation}
A microscopic description of the LLG equation can be obtained by treating the phonon system as a reservoir which influences the magnon dynamics. This is accomplished by use of the interaction picture master equation \cite{breuer_theory_2002}.
\begin{align}\label{eq:mastereq}
\frac{d}{dt}\rho_m=-tr_R\left\{\int_0^t\left[H_I(t),\left[H_I(s),\rho_m(s)\otimes\rho_R\right]\right]ds\right\}.
\end{align}
Where $\rho_m=tr_R(\rho)$ is the magnon system density matrix, $\rho_R$ is the phonon reservoir density matrix and $\rho$ the global density matrix. Above we assume the Born approximation in which $\rho(t)\approx\rho_m(t)\otimes\rho_R$ at all times. This is consistent with a truncation of the Nakajima-Zwanzig memory kernel at quadratic order in the interaction strength $V$ \cite{breuer_theory_2002,grabert_projection_1982}. Critically, \prettyref{eq:mastereq} is \textit{non-Markovian}. That is, the evolution of the magnon system at time $t$ is dependent on its state at all prior times $s<t$.

Expectation values then evolve in time according to
\begin{align}
\frac{d}{dt}\braket{A}=tr_R\of{\rho_m\frac{\partial A}{\partial t}}+tr\of{\frac{d\rho_m}{dt}A}.
\end{align}
The first term concerns purely magnon system dynamics, and is discussed in the previous section. The remaining dynamics are described by the master equation. Using this strategy we find an additional longitudinal relaxation for the magnetization $\frac{d}{dt}\mathbf{M}\sim-\mathbf{M}$ stemming from a loss of coherence of the spin system.

If the reservoir is fully incoherent it may be treated as an Ohmic reservoir, and we obtain a damping from the phonon reservoir $\frac{d}{dt}\mathbf{M}\sim\mathbf{M}\times\frac{d}{dt}\mathbf{M}$. Below we show that this treatment is appropriate in the low frequency limit where coherent excitations in the reservoir relax much faster than the dynamics of the spin system. At higher frequencies we find an additional nutation term $\frac{d}{dt}\mathbf{M}\sim\mathbf{M}\times\frac{d^2}{dt^2}\mathbf{M}$. The nutational dynamics are then interpreted as a coherent interaction with the phonon reservoir- the spin systems interacts with a phonon at time $t$ and then again at time $t'$ giving the magnetization inertia. On timescales much longer than the phonon lifetime only the longitudinal decay and Gilbert damping are present. 

\subsection{Master Equation}
Using \prettyref{eq: vertex} for the interaction Hamiltonian in \prettyref{eq:mastereq} we find for the evolution of an operator $A$ within the magnon Hilbert space
\begin{align} \label{eq: dA}
tr\Big(\frac{d\rho_m}{dt}A\Big)&=\int_0^t    tr_m\{tr_R(\left[H_I(t),\left[H_I(s),\rho_m(s)\otimes\rho_R\right]\right])A(t)\}ds=\nonumber\\
    &=\SUM{kk'}{}\sum_{\lambda\lambda'}V^*_{k,\lambda}V^*_{k',\lambda'}{}\of{\braket{[A,a_{k'}]a_k}_m\braket{[B_{k',\lambda'}^{\dagger},B_{k,\lambda}^{\dagger}]}_R+\braket{[[A,a_{k'}],a_k]}_m\braket{B_{k,\lambda}^{\dagger}B_{k',\lambda'}^{\dagger}}_R}+\nonumber\\
    &+\SUM{kk'}{}\sum_{\lambda\lambda'}V^*_{k,\lambda}V_{k',\lambda'}\of{\braket{[A,a_{k'}^{\dagger}],a_k}_m\braket{[B_{k',\lambda'},B_{k,\lambda}^{\dagger}]}_R+\braket{[[A,a_{k'}^{\dagger}],a_k]}_m\braket{B_{k,\lambda}^{\dagger}B_{k',\lambda'}}_R}+\nonumber\\
        &+\SUM{kk'}{}\sum_{\lambda\lambda'}V_{k,\lambda}V^*_{k',\lambda'}\of{\braket{[A,a_{k'}],a_k^{\dagger}}_m\braket{[B_{k',\lambda'}^{\dagger},B_{k,\lambda}]}_R+\braket{[[A,a_{k'}],a_k^{\dagger}]}_m\braket{B_{k,\lambda}B_{k',\lambda'}^{\dagger}}_R}+\nonumber\\
            &+\SUM{kk'}{}\sum_{\lambda\lambda'}V_{k,\lambda}V_{k',\lambda'}\of{\braket{[A,a_{k'}^{\dagger}]a_k^{\dagger}}_m\braket{[B_{k',\lambda'},B_{k,\lambda}]}_R+\braket{[[A,a_{k'}^{\dagger}],a_k^{\dagger}]}_m\braket{B_{k,\lambda}B_{k',\lambda'}}_R}.
\end{align}
Where $k'$ accompanies time $t$ and $k$ accompanies $s$ in each term. The subscripts on the brackets denote the subspace in which each expectation value is evaluated, $\braket{\dots}_i=tr_i(\rho_i\dots)$. For $A(t)=a_p(t)$
\begin{align}
    \braket{\dot{\rho}_ma_p(t)}&=-\SUM{k}{}\sum_{\lambda,\lambda'}\int_0^tds\Big\{\braket{a_k(s)}_mV_{p,\lambda'}V_{k,\lambda}^*\braket{[B_{p\lambda'},B_{k,\lambda}^{\dagger}]}_R+\braket{a_k^{\dagger}(s)}_mV_{p,\lambda'}V_{k,\lambda}\braket{[B_{p,\lambda'},B_{k,\lambda}]}_R\Big\}\nonumber\\
    &=-i\sum_k\int_0^{\infty}ds\Big\{\mathcal{K}_{p,k}(t-s)\braket{a_k(s)}_m+\mathcal{J}_{p,k}(t-s)\braket{a^{\dagger}_k(s)}_m\Big\}.
\end{align}
Where we have identified the memory kernels
\begin{align}\label{eq: K}
   &\mathcal{K}_{p,k}(t-s)=-i\sum_{\lambda,\lambda'}V_{p,\lambda'}V_{k,\lambda}^*\theta(t-s)\braket{[B_{p,\lambda'}(t),B_{k,\lambda}^{\dagger}(s)]}_R\\\label{eq: J}
   &\mathcal{J}_{p,k}(t-s)=-i\sum_{\lambda,\lambda'}V_{p,\lambda'}V_{k,\lambda}\theta(t-s)\braket{[B_{p,\lambda'}(t),B_{k,\lambda}(s)]}_R   
\end{align}

The reservoir correlation functions $\braket{B_p(t)B^{\dagger}_k(s)}_R=tr_R\of{\rho_RB_{p}(t)B^{\dagger}_k(s)}$ can be equivalently considered within the Heisenberg picture for the isolated reservoir. The memory kernel we therefore recognize as being similar to the Born correction to the magnon self energy, but with the full retarded Green's function of the phonon system $\mathcal{K}_{p,k}(t-s)=V_pV^*_kD^R_{p,k}(t-s)$.

The dynamics of the reservoir will excite magnon modes $k\neq p$. In terms of the magnetization dynamics, this means that there will be an additional decay in the amplitude of $\mathbf{M}$ from the coupling to the higher $k$ modes. Let $\mathcal{K}_0=\mathcal{K}_{0,0}$, $\mathcal{J}_{0}=\mathcal{J}_{0,0}$, and $\Gamma=-iM_S\sqrt{\frac{2}{NS}}\SUM{k\neq0}{}\of{\mathcal{K}_{0,k}*\braket{a_k}+\mathcal{J}_{0,k}*\braket{a^{\dagger}_k}}$, where $(f*g)(t)=\int_0^{\infty}dsf(t-s)g(s)$ denotes convolution in time. Focusing for the moment on the portion of the dynamics coming from the reservoir we define $\dbar\braket{A}=tr(\dot{\rho}_mA)$ The transverse components then evolve as
\begin{align}
    &\dbar M_x=Im(\mathcal{J}_0 * M_x) + \mathcal{K}_0 * M_y - Re(\mathcal{J}_0 * M_y) + Re(\Gamma)\\
    &\dbar M_y = -Im(\mathcal{J}_0 * M_y) -\mathcal{K}_0 * M_x - Re(\mathcal{J}_0 * M_x) + Im(\Gamma).
\end{align}
Note that at zero momentum the phonon Green's function is real, and so $\mathcal{K}_0=\overline{\mathcal{K}}_0$. The decay term scales with the off-diagonal Green's function $\Gamma\sim D^R_{0,k}(t-s)$. If the reservoir can be considered translationally invariant then these terms vanish due to conservation of momentum. In the presence of disorder, impurities or domain and sample boundaries one has more generally that $D^R_{0,k}\sim D_0^R\Pi_{0,k}D_k^R$, where $\Pi_{0,k}$  is the self energy for the non-momentum conserving scattering process. $\Gamma$ is suppressed by an extra Green's function relative to $\mathcal{K}_0$ and $\mathcal{J}_0$, and will therefore have a negligible influence on the dynamics. Thus, we approximate $\Gamma\approx0$ in what follows.

At zero momentum the phonon polarization vectors may be chosen to be real. Then
\begin{align}
    \dbar \mathbf{M}=-2S^3z^2\sum_{\lambda\lambda'}(\nabla K\cdot d_{0,\lambda})(\nabla K\cdot d_{0,\lambda'})W^TD_{0,\lambda',\lambda}^R*\mathbf{M}
\end{align}
where the matrix $W=\mathbf{n}\mathbf{n}^TX$ is described in the previous section. Define the transverse memory kernel 
\begin{align}
    \mathcal{T}(t-s)=2S^3z^2\sum_{\lambda,\lambda'}(\nabla K\cdot d_{0,\lambda})(\nabla K\cdot d_{0,\lambda'})D_{0,\lambda',\lambda}^R(t-s).
\end{align}
Including the system dynamics in \prettyref{eq: ellipse} we find for the transverse components of the magnetization
\begin{align} \label{eq: transverse}
    \dot{\mathbf{M}}=(\omega_0+KSn^2)\mathbf{M}\times\hat{z}+2KSn^2(\hat{\mathbf{n}}\hat{\mathbf{n}}^T-1)(\mathbf{M}\times\hat{z})-n^2(\hat{\mathbf{n}}\hat{\mathbf{n}}^T-1)(\mathcal{T}*\mathbf{M}\times\hat{z}).
\end{align}
Again the dynamics are elliptical. The memory kernel is responsible for both damping and nutation at high frequencies. In the limit of an ohmic reservoir (described in detail in the following section) $\mathcal{T}(\omega)\sim -i\kappa\omega$ with $\kappa>0$ a real parameter describing the strength of the damping. The solution to \prettyref{eq: transverse} is then an elliptical spiral towards the equilibrium magnetization, \prettyref{fig: spiral}. As discussed in the previous section the ellipse may be treated as approximately circular for strong external fields, $\omega_0>>KS.$ Neglecting the projection operators we find a natural extension to the full, nonlinear $LLG$ equation.
\begin{align} \label{eq: TLLG}
    \dot{\mathbf{M}}=-\gamma\mathbf{M}\times\mathbf{H}_{eff}+\frac{n^2}{M_s}\mathbf{M}\times\mathcal{T}*\mathbf{M}+\frac{1}{2M_S^2}\frac{d}{dt}(M^2)\mathbf{M}.
\end{align}
$\mathbf{H}_{eff}=(H+2KSz^2/|\gamma|)\hat{z}$ is the effective field and we have used that $\mathbf{M}\times\mathcal{T}*\mathbf{M}\approx M_S\hat{z}\times\mathcal{T}*\mathbf{M}$ and $\frac{\mathbf{M}\cdot\dot{\mathbf{M}}}{M_S^2}\mathbf{M}\approx\frac{\mathbf{M}\cdot\dot{\mathbf{M}}}{M_S}$ within the spin wave approximation. The last term has been included to make this extension consistent with \prettyref{eq: Mz}. The spin wave approximation leads to a linearized approximation to the full LLG equation in which the components of $\mathbf{M}$ transverse to the effective field evolve independently of $M_z$, \prettyref{eq: transverse}. This is no longer the case in the extension to \prettyref{eq: TLLG}, and as a consequence the magnitude of $\mathbf{M}$ is not preserved. $\frac{d}{dt}(M^2)$ may be calculated from \cref{eq: Msquared,eq: dA}, though its solution is beyond the scope of this article.
\begin{comment}
Let $A(t)=\sum_{p\neq0}a^{\dagger}_p(t)a_p(t)$,
\begin{align}
    d\braket{A}=&\sum_{p\neq0}\sum_{k}\Big( -i\mathcal{J}^*_{p,k}(t-s)*\braket{a_{p}(t)a_k(s)}+i\mathcal{K}_{p,k}(t-s) * \braket{a^{\dagger}_{p}(t)a_k(s)}-i\mathcal{K}_{p,k}^* * \braket{a_{p}(t)a_k^{\dagger}(s)}+i\mathcal{J}_{p,k} * \braket{a^{\dagger}_{p}(t)a^{\dagger}_k(s)}\Big)+\\
    +&\sum_{p\neq0}\sum_k\Big(-i\mathcal{F}_{p,k}\braket{[a_{p}(t),a_k(s)]}+
\end{align}
\end{comment}

The memory kernel is directly proportional to the retarded phonon Green's function at zero frequency. For acoustic modes $D^R_{k=0}=0$, suggesting that only optical phonon modes couple to the transverse magnetization. In the following section we provide a quasiparticle description of this Green's function from which we interpret damping as arising from decoherence within the reservoir, and nutation as a consequence of coupling to the high frequency optical phonon mode.

\begin{figure}[tph]
\centering
 \begin{tikzpicture}[scale=1.25]
    \def\a{1}
    \def\b{0.1}
    \def\g{-0.1}
    
    \draw[->] (0,0) -- (0,1.5) node[right] {$y$};
    \draw[->] (0,0) -- (1.5,0) node[right] {$x$};
    \draw[thick,red] plot[domain=0:10*pi,samples=100,smooth] 
        ({e^(\g*\x)*(-\g/\a*sin(\a*\x r)+cos(\a*\x r))},{e^(\g*\x )*(-\a)/(\a+\b)*sin(\a*\x r)});
    \draw[thick,blue] plot[domain=0:10*pi,samples=100,smooth] 
        ({(-\g/\a*sin(\a*\x r)+cos(\a*\x r))},{(-\a)/(\a+\b)*sin(\a*\x r)});    
    \end{tikzpicture}
    \caption{\label{fig: spiral} Solution to \prettyref{eq: transverse} in the ohmic limit with $2KS=0.1\omega_0$ and damping $\kappa\omega_0=0.1$. The blue curve shows the undamped elliptical orbit. Damping causes an elliptical spiral towards the equilibrium $M_x=M_y=0$.}
    \end{figure}
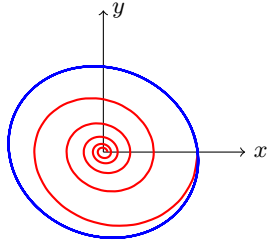

\section{Phonon Green's Function}
In what follows we assume the phonon bands are non-degenerate at zero momentum so that $D^R_{0,\lambda,\lambda'}\sim\delta_{\lambda,\lambda'}$. In general the degeneracy of the phonon branches at the $\Gamma$ point are determined by the dimensions of the irreducible representations of the point group of the crystal. The arguments presented here are then applicable to those representations with dimension one.

For a single, non-degenerate branch the retarded single particle phonon Greens function is\footnote{We have dropped the supersript $R$ to simplify the notation}
\begin{align} \label{eq: Dq}
    D_q(\omega)=\frac{2\omega_q}{\omega^2-\omega_q^2-2\omega_q\Pi_q(\omega)}
\end{align}
With $\Pi(\omega)$ the reservoir self energy, which excludes the magnon-phonon interaction. Define the complex frequency $\Omega_q^2(\omega)=\omega_q^2-2\omega_q\Pi(\omega)$. \prettyref{eq: Dq} can then be decomposed as
\begin{align}\label{eq: Dsep}
    D_q(\omega)=\frac{\omega_q}{\Omega_q}\Big[\frac{1}{\omega-\Omega_q}-\frac{1}{\omega+\Omega_q}\Big].
\end{align}
Defining $\Omega=\nu-i\Gamma$, the Green's function will have poles at the real frequencies $\omega_{\pm}=\pm\nu(\omega_{\pm}).$ As we proceed we drop the subscript $q$ from $\Omega$ and use the shorthand $\nu(\omega_{\pm})=\nu_{\pm}$ and $\Gamma(\omega_{\pm})=\Gamma_{\pm}.$ Assuming the Green's function is analytic about either pole we may expand $\Omega=\Omega_++\Omega'_+(\omega-\omega_+)+\pi_+(\omega)$, where $\pi_+(\omega)\sim(\omega-\omega_+)^2$ as $\omega\rightarrow\omega_+$. The denominator in the first term of \prettyref{eq: Dsep} can then be written.
\begin{align}\label{eq: expand}
    \omega-\Omega=(\omega-\omega_+)(1-\nu'_+)+i\Gamma_++i\Gamma'_+(\omega-\omega_+)-\pi_+(\omega).
\end{align}
Define now $\sigma_+(\omega)=\frac{Z_+\Omega}{\omega_+}[-i\Gamma'_+(\omega-\omega_+)+\pi_+(\omega)]-\frac{\Omega-\omega_+}{\omega_+}[(\omega-\omega_+)+iZ_+\Gamma_+]$ and $Z_+=(1-\nu_+')^{-1}$. The first term in \prettyref{eq: Dsep} then takes the form
\begin{align} \label{eq: first}
    \frac{\omega_q}{\Omega}\frac{1}{\omega-\Omega}=\frac{\omega_q}{\omega_+}\frac{Z_+}{\omega-\omega_++iZ_+\Gamma_+-\sigma_+(\omega)}.
\end{align}

At the pole, $-\sigma_+(\omega_+)=Z_+\Gamma_+^2/\omega_+<<1$. If $D_q(\omega)$ has only simple, isolated poles at $\omega_{\pm}$ then $\sigma_+(\omega)$ will have no other roots. The pole structure of the Green's function is therefore captured by
\begin{align} \label{eq: D+}
    D_+(\omega)=\frac{\omega_q}{\omega_+}\frac{Z_+}{\omega-\omega_+
    +iZ_+\Gamma_+}
\end{align}
which describes a quasi-particle excitation at $\omega=\omega_+$ with lifetime $\tau_+=1/(Z_+\Gamma_+)$. The difference between \prettyref{eq: first} and \prettyref{eq: D+} describes an incoherent\footnote{Incoherent because it has not distinct pole structure.} background
\begin{align}
    D_+^{inc}(\omega)&=D_+(\omega)\frac{\beta_+(\omega)}{1-\beta_+(\omega)},\\\nonumber\\
    \beta_+(\omega)&=\frac{\sigma(\omega)}{\omega-\omega_++iZ_+\Gamma_+}.
\end{align}
At the pole $D_+^{inc}(\omega_+)=\frac{\omega_q}{\omega_+}\frac{1}{\omega_+-i\Gamma_+}\sim1/\omega_+<<|D_+(\omega_+)|$. By definition $\beta_+(\omega)\neq1$ and so far from the pole the incoherent Green's function will be slowly varying. Asymptotically $D_+(\omega)$ decays as $1/|\omega-\omega_+|$, and therefore $D_+^{inc}\sim D$ for $|\omega-\omega_+|>>Z_+\Gamma_+$.

Although this expansion \prettyref{eq: expand} was carried out near the pole, this has only been used to specify the limiting behavior of the function $\pi_+(\omega)$ near $\omega\sim\omega_+$, which is otherwise unspecified. We stress, therefore, that the decomposition $D=D_++D_+^{inc}$ holds for all $\omega$, provided that the phonon self energy is well behaved. 

This procedure can be replicated near conjugate pole for the second term in \prettyref{eq: Dsep}, from which we find
\begin{align}
    D_-(\omega)&=-\frac{\omega_q}{\omega_-}\frac{Z_-}{\omega-\omega_--iZ_-\Gamma_-}\\\nonumber\\
    Z_-&=(1+\nu_-')^{-1}\\\nonumber\\
    D_{-}^{inc}(\omega)&=D_-(\omega)\frac{\beta_-(\omega)}{1-\beta_-(\omega)}\\\nonumber\\
    \beta_-(\omega)&=\frac{\sigma_-(\omega)}{\omega-\omega_--iZ_-\Gamma_-}\\\nonumber\\
    \sigma_-(\omega)&=\frac{Z_-\Omega}{\omega_-}[-i\Gamma'_-(\omega-\omega_-)+\pi_-(\omega)]+\frac{\Omega+\omega_-}{\omega_-}[(\omega-\omega_-)-iZ_-\Gamma_-].
\end{align}
Finally, we have for the full phonon Green's function
\begin{align}\label{eq: Ddecomp}
    D_q(\omega)&=D_+(\omega)-D_-(\omega)+D^{inc}(\omega)\\\nonumber\\
    D^{inc}(\omega)&=D_+^{inc}(\omega)-D_-^{inc}(\omega).
\end{align}

In order for the quasi-particles to be well defined their spectral function $A_{\pm}(\omega)=-\frac{1}{\pi}Im(D_{\pm}(\omega))$ must be reasonably narrow, which requires $Z_{\pm}\Gamma_{\pm}<<|\omega_{\pm}|$. This will hold if the phonon self energy is small compared to the bare frequency $\omega_q$. In this case $\Omega_q\approx\omega_q+\Pi_q(\omega)$ and $\nu-i\Gamma=Re(\Pi)+iIm(\Pi)$. The residues for the quasi-particle Green's functions are therefore given in the usual way
\begin{align}
    Z_{\pm}=(1\mp\partial_{\omega}Re\Pi_q(\omega_{\pm}))^{-1},
\end{align}
with broadening
\begin{align}
    \Gamma_{\pm}=-Im\Pi_q(\omega_{\pm}).
\end{align}
Of particular interest for this work are the $\mathbf{q}=0$ modes. In the time domain $D_0(t-s)=D_0^*(t-s)$ is real, which enforces parity restrictions on $\Pi_0(\omega)$, namely $Re(\Pi_0)$ is even and $Im(\Pi_0)$ is odd. In this case $\omega_-=-\omega_+$, $Z_-=Z_+$ and $\Gamma_-=-\Gamma_+$. If the shift in frequency is reasonably small then we can approximate $\omega_{\pm}\approx\pm\Omega_0,$ with $\Omega_0$ the frequency of an optical phonon at zero momentum. Then
\begin{align}
    D_{\pm}=\frac{Z}{\omega\mp\Omega_0+iZ\Gamma}.
\end{align}
Because $ImD_0(\omega)$ must be odd, at leading order $ImD^{inc}(\omega)\sim\omega$ as $\omega\rightarrow0.$ From \prettyref{eq: TLLG} we note that the constant contribution to the Green's function is irrelevant for the magnetization dynamics, and so we estimate the incoherent contribution as $D^{inc}(\omega)\approx -i\kappa\omega$. Assuming the reservoir remains in equilibrium, the constant $\kappa$ may be fixed by the normalization condition for the spectral function \cite{mahan_many-particle_2000}
\begin{align}
    -\frac{1}{\pi}\int_{-\infty}^{\infty}d\omega ~n_B(\omega)ImD^R_0(\omega)=2n_0+1.
\end{align}
$n_B(\omega)$ is the Bose-Einstein distribution, and $n_0$ is the occupation of the optical mode. Integrating along a contour in the upper half of the complex plane
\begin{align}
    -\frac{1}{\pi}\int d\omega~n_B(\omega)Im(D_+-D_-)=Z[n_B(\Omega_0+iZ\Gamma)-n_B(-\Omega_0+iZ\Gamma)]-2iZ\sum_{n=1}^{\infty}Im[D_+(i\omega_n)-D_-(i\omega_n)].
\end{align}
For the imaginary Matsubara frequencies $i\omega_n=2\pi i n/\beta$ with integer $n$ and inverse temperature $\beta$, $D_+(i\omega_n)^*=-D_-(i\omega_n)$. Each term in the Matsubara sum above therefore vanishes because $D_+(i\omega_n)-D_-(i\omega_n)$ is real. Up to second order in $Z\Gamma/\Omega_0$, $n_B(\Omega_0+iZ\Gamma)-n_B(-\Omega_0+iZ\Gamma)=2n_0+1-(Z\Gamma)^2n''_0$, where we have used that $-n_B(-\omega)=n_B(\omega)+1$. With this identity we see that the linear order term vanishes because $n_B'$ is odd, and the second order term may be written $n''_0=-\beta n'_0(1+2n_0)$. It is interesting to interpret the derivative $n'_0$ as fluctuations in the occupation due to broadening of the optical mode, $Z\Gamma n'_0=\delta n_0$. From \prettyref{eq: Ddecomp}, then,
\begin{align}\label{eq: constraint}
    -\frac{1}{\pi}\int d\omega~n_B(\omega)ImD^{inc}(\omega)=(2n_0+1)[1-Z(1+\beta Z\Gamma\delta n_0)].
\end{align}
Assuming a cutoff $D^{inc}\rightarrow0$ for $|\omega|>\Omega_0$ we find for the damping parameter
\begin{align}
    \kappa\approx\frac{2\pi}{\Omega_0^2}(2n_0+1)[1-Z(1+\beta Z\Gamma\delta n_0)].
\end{align}
Note that the analogy we have drawn with Fermi liquid theory is not exact in that the constraint \prettyref{eq: constraint} does not force $0<Z<1$ for the quasiparticle residue. However, this is still approximately satisfied since $Z\Gamma/\Omega_0<<1$, and so $Z$ provides an approximate measure of the degree of coherence of the optical mode.

The incoherent background in the phonon reservoir therefore provides an ohmic contribution to the memory kernel at small frequencies similar to what is assumed in the Caldeira-Leggett model used in \cite{quarenta_bath-induced_2024}. Here, however, the transverse dynamics couple primarily to a single $k=0$ optical mode, and the damping arises from broadening of this mode within the reservoir. We stress that this is distinct from the usual assumptions of system reservoir interactions, in which the system modes are broadened by a continuum of reservoir modes.

\end{onecolumngrid}
\bibliography{iLLG_refs}